\documentclass[twocolumn,prb,showpacs,floatfix]{revtex4}
\usepackage{amssymb}
\usepackage{amsmath}
\usepackage{acronym}
\usepackage{amstext}
\usepackage{graphicx}

\begin{document}

\title{The current phase relation in Josephson tunnel junctions.}
\author{A.~A.~Golubov}
\email{a.golubov@tn.utwente.nl}
\affiliation{Faculty of Science and Technology, University of Twente, The Netherlands}
\author{M.~Yu.~Kupriyanov}
\email{mkupr@pn.sinp.msu.ru}
\affiliation{Nuclear Physics Institute, Moscow State University, 119992 Moscow, Russia}
\date{7 Febraury 2005}

\begin{abstract}
The $J(\varphi )$ relation in SFIFS, SNINS and SIS tunnel junctions
is studied. The method for analytical solution of linearized Usadel
equations has been developed and applied to these structures. It is
shown that the
Josephson current across the structure has the sum of $\sin \varphi $ and $%
\sin 2\varphi $ components. Two different physical mechanisms are
responsible for the sign of $\sin 2\varphi $. The first one is the
depairing by current which contributes positively to the $\sin
2\varphi $ term, while the second one is the finite transparency of
SF or SN interfaces which provides the negative contribution. In
SFIFS junctions, where the first harmonic vanishes at $"0"$ - $"\pi
"$ transition, the calculated second harmonic fully determines the
$J(\varphi )$ curve.
\end{abstract}

\pacs{74.50.+r, 74.80.Dm, 75.30.Et}
\maketitle

It is well known that tunnel SIS Josephson junctions have sinusoidal
current-phase relation, while with the decrease of the barrier transparency
deviations from $\sin \varphi $ take place (see \cite{Likharev,GKI} for the
review). The sign of second harmonic is important for many applications, in
particular in junctions with a more complex structure like SNINS or SFIFS,
where N is a normal metal and F is a weak metallic ferromagnet \cite%
{Ryazanov,Frolov,GKI}. To analyze \ this problem selfconsistently, one
should go beyond the approximation which is usually used and is called
"Rigid boundary conditions" (RBC) .

The \ RBC method is an effective tool extensively used earlier for
theoretical study of the proximity and Josephson effects \cite{Likharev,GKI}%
. This method is based on the assumption that all nonlinear and
nonequlibrium effects in a Josephson structure are located in a "weak link"
connecting two superconducting electrodes. The back influence of these
effects on superconductivity in the\ electrodes is neglected. The RBC are
valid if a junction has the constriction geometry. The quantitative criteria
for the validity of RBC for planar SIS tunnel junctions, SS$^{\prime }$S
sandwiches and variable thickness bridges were studied only numerically for
some parameter ranges\cite{GKI}. The main technical difficulty in
formulating the analytical criteria of RBC validity is to find the solution
of equations describing the perturbation of superconducting state in S
electrodes. In this paper we will attack this problem by finding the
solution of linearized Usadel equations \cite{Usadel}. We will also use this
solution to formulate the corrections to previous results obtained in RBC\
approximation.

\textbf{The junction model.} Let us consider the structure of SFIFS type,
where for simplicity the parameters of the SF bilayers are equal to each
other. We assume that the S layers are bulk and that the dirty limit
conditions are fulfilled in the S and F metals. We assume further that F
metals are weak monodomain ferromagnets with zero electron--phonon
interaction constant and the FS interfaces are not magnetically active. We
will restrict ourselves to the case of parallel orientation of the exchange
fields $H$ in the ferromagnets. The results obtained for SFIFS junctions
cross over to SNINS and SIS in corresponding limits.

Under the above assumptions the problem is reduced to the solution of the
one-dimensional Usadel equations \cite{Usadel,Buzdin1} in S- and F-layers
and matching these solutions by the appropriate boundary conditions.\cite{KL}
We choose the $x$ axis perpendicular to the plane of the interfaces with the
origin at the central barrier I and introduce indexes $L$ (left), $R$
(right) and $I$ for description the materials and interfaces parameters of
the SFIFS structure located on the left and right sides from the central
barrier and at this central barrier, respectively.

The Usadel functions $G$ and $F$ obey the normalization condition $G_{\omega
}^{2}+F_{\omega }F_{-\omega }^{\ast }=1$, which allows the following
parametrization in terms of the new function $\Phi $:
\begin{equation}
G_{\omega }=\frac{\widetilde{\omega }}{\sqrt{\widetilde{\omega }^{2}+\Phi
_{\omega }\Phi _{-\omega }^{\ast }}},\quad F_{\omega }=\frac{\Phi _{\omega }%
}{\sqrt{\widetilde{\omega }^{2}+\Phi _{\omega }\Phi _{-\omega }^{\ast }}}.
\label{def_f}
\end{equation}%
The quantity $\widetilde{\omega }=\omega +iH$ corresponds to the general
case when the exchange field $H$ is present. However, in the S layers $H=0$
and we have simply $\widetilde{\omega }=\omega $.

The Usadel equations \cite{Usadel} in the S and F layers have the form
\begin{gather}
\xi _{S}^{2}\frac{\pi T_{c}}{\omega G_{S}}\frac{\partial }{\partial x}\left[
G_{S}^{2}\frac{\partial }{\partial x}\Phi _{S}\right] -\Phi _{S}=-\Delta ,
\label{EqUS} \\
\xi _{F}^{2}\frac{\pi T_{c}}{\widetilde{\omega }G_{F}}\frac{\partial }{%
\partial x}\left[ G_{F}^{2}\frac{\partial }{\partial x}\Phi _{F}\right]
-\Phi _{F}=0,  \label{EqUSF}
\end{gather}%
where $G_{\omega }=\widetilde{\omega }/\sqrt{\widetilde{\omega }^{2}+\Phi
_{\omega }\Phi _{-\omega }^{\ast }}$, $\widetilde{\omega }=\omega +iH$ in a
ferromagnet ($H$ is the exchange field), $\widetilde{\omega }=\omega $ in S
and N metals, $T_{c}$ and $\Delta $ are the critical temperature and the
pair potential in a superconductor, $\omega =\pi T(2n+1)$ are the Matsubara
frequencies and $\xi _{S(F)}$ are the coherence lengths related to the
diffusion constants $D_{S(F)}$ as $\xi _{S(F)}=\sqrt{D_{S(F)}/2\pi T_{c}}$.
The pair potential satisfies the self-consistency equations
\begin{equation}
\Delta \ln \frac{T}{T_{c}}+\pi T\sum_{\omega =-\infty }^{\infty }\frac{%
\Delta -G_{S}\Phi _{S}sgn\omega }{|\omega |}=0.  \label{EqDEL}
\end{equation}

In the case of SFIFS tunnel junction in quasi-one dimensional geometry the
boundary conditions at the junction plane ($x=0$) read
\begin{gather}
\xi _{F}\frac{G_{F,L}^{2}}{\widetilde{\omega }_{L}}\frac{\partial }{\partial
x}\Phi _{F,L}=\xi _{F}\frac{G_{F,R}^{2}}{\widetilde{\omega }_{R}}\frac{%
\partial }{\partial x}\Phi _{F,R},  \label{BI_1} \\
\gamma _{BI}\frac{\xi _{F}G_{FL,R}}{\widetilde{\omega }_{L}}\frac{\partial }{%
\partial x}\Phi _{FL,R}=\pm G_{F,R}\left( \frac{\Phi _{F,R}}{\widetilde{%
\omega }_{R}}-\frac{\Phi _{F,L}}{\widetilde{\omega }_{L}}\right) ,
\label{BI_2} \\
\text{with}\quad \gamma _{BI}=R_{N}\mathcal{A}_{I}/\rho _{F}\xi _{F},  \notag
\end{gather}%
where the indices $L$ and $R$ refer to the left- and right-hand side of the
junction, respectively, $R_{N}$ and $\mathcal{A}_{I}$ are the normal
resistance and the area of FIF interface.

The boundary conditions at the SF interfaces ($x=\mp d_{F}$) have the form
\cite{KL}
\begin{gather}
\frac{\xi _{S}G_{S,k}^{2}}{\omega }\frac{\partial }{\partial x}\Phi
_{S,k}=\gamma \frac{\xi _{F}G_{F,k}^{2}}{\widetilde{\omega }_{k}}\frac{%
\partial }{\partial x}\Phi _{F,k},  \label{BC_Fi1} \\
\pm \gamma _{B}\frac{\xi _{F}G_{F,k}}{\widetilde{\omega }_{k}}\frac{\partial
}{\partial x}\Phi _{F,k}=G_{S,k}\left( \frac{\Phi _{S,k}}{\omega }-\frac{%
\Phi _{F,k}}{\widetilde{\omega }_{k}}\right) ,  \label{BC_Fi2} \\
\text{with}\quad \gamma _{B}=R_{B}\mathcal{A}_{B}/\rho _{F}\xi _{F},\quad
\gamma =\rho _{S}\xi _{S}/\rho _{F}\xi _{F},  \notag
\end{gather}%
where $R_{B}$ and $\mathcal{A}_{B}$ are the resistance and the area of the
SF interfaces; $\rho _{S(F)}$ is the resistivity of the S (F) layer; $k=L,R.$%
. Both of these conditions ensure continuity of the supercurrent.

We will also suppose that due to low transparency of the FIF interface the
Josephson current is much smaller that the depairing current of
superconducting electrodes so that the suppression of superconductivity in
the interior of the electrodes can be neglected and at $x\rightarrow \pm
\infty $%
\begin{equation}
\left\vert \Phi _{S,k}\right\vert =\Delta _{0},  \label{BC_inf}
\end{equation}%
where $\Delta _{0}$ is the magnitude of bulk order parameter.

\textbf{\ The limit of small F layer thickness}. In this limit
\begin{equation}
d_{F}\ll \min \left( \xi _{F},\sqrt{\frac{D_{F}}{2H}}\right)
\label{cond_small}
\end{equation}%
the gradients in (\ref{EqUSF}) are small and in the second approximation on $%
d_{F}/\xi _{F}$ the solution of (\ref{EqUSF}) has the form
\begin{equation}
\Phi _{F,k}=A_{k}+B_{k}\frac{x}{\xi _{F}}+\frac{x^{2}}{2}\frac{\widetilde{%
\omega }_{k}A_{k}}{\pi T_{c}\xi _{F}^{2}G_{F,k}},  \label{Sol_1_pr}
\end{equation}%
\begin{equation*}
G_{F,k}^{2}=\frac{\widetilde{\omega }_{R}^{2}}{\widetilde{\omega }%
_{R}^{2}+A_{k}^{2}(\omega )}.
\end{equation*}%
Integration constants $\tilde{A}$ and $\tilde{B}$ in (\ref{Sol_1_pr}) can be
found from boundary conditions at $x=0$%
\begin{equation}
\frac{G_{F,L}^{2}}{\widetilde{\omega }_{L}}B_{L}=\frac{G_{F,R}^{2}}{%
\widetilde{\omega }_{R}}B_{R}=\frac{G_{F,L}G_{F,R}}{\gamma _{BI}}\left(
\frac{A_{R}}{\widetilde{\omega }_{R}}-\frac{A_{L}}{\widetilde{\omega }_{L}}%
\right)  \label{bcBLR}
\end{equation}%
and at $x=\pm d_{F}$

\begin{equation}
A_{k}=A_{0,k}\mp \gamma _{B}\frac{G_{F,k}}{G_{S,k}+\widetilde{\omega }%
_{k}\gamma _{BM}/\pi T_{c}}B_{k},  \label{GammaBM}
\end{equation}%
\begin{equation}
A_{0,k}=\frac{\widetilde{\omega }_{R,L}\Phi _{S,k}G_{S,k}}{\omega \left(
G_{S,k}+\widetilde{\omega }_{k}\gamma _{BM}/\pi T_{c}\right) },\quad \gamma
_{BM}=\gamma _{B}\frac{d_{F}}{\xi _{F}}.  \label{Arov}
\end{equation}%
Expression (\ref{GammaBM}) valid if $\gamma _{B}\ll \gamma _{BI}$.
Substitution of (\ref{Sol_1_pr}) and (\ref{GammaBM}) into the boundary
condition at $x=\pm d_{F}$ leads to

\begin{equation}
\xi _{S}\frac{\partial }{\partial x}\Phi _{S,k}=\pm \gamma _{M}\frac{G_{F,k}%
}{G_{S,k}^{2}}\frac{\omega }{\pi T_{c}}A_{k}+\gamma \frac{\omega G_{F,k}^{2}%
}{\widetilde{\omega }_{k}G_{S,k}^{2}}B_{k},  \label{GammM}
\end{equation}%
where $\gamma _{M}=\gamma d_{F}/\xi _{F}$ and reduce boundary problem (\ref%
{EqUS})--(\ref{BC_inf}) to the solution of equations (\ref{EqUS}), (\ref%
{EqDEL}) in the S-layers with the boundary conditions (\ref{BC_inf}), (\ref%
{GammM}). At $H=0$ and ($\gamma _{BI}d/\xi _{F})\gg 1$ expression ( \ref%
{GammM}) reduces to the known result for SN bilayer. \cite{GKJ}

\textbf{The linearized Usadel equations}. Following RBC\ approximation we
will start with the assumption that the suppression of superconductivity in
S layer is weak and the solution of Usadel equations in the superconductor
has the form
\begin{eqnarray}
\Phi _{S,k}(\omega ) &=&\Delta _{0,k}+\Phi _{1,k},\quad \Delta =\Delta
_{0,k}+\Delta _{1,k},  \label{rigid} \\
G_{S,k} &=&G_{0}+G_{1,k},\quad G_{0}=\frac{\omega }{\sqrt{\omega ^{2}+\Delta
_{0}^{2}}} \\
G_{1,k} &=&-\frac{G_{0}}{\omega ^{2}+\Delta _{0}^{2}}\frac{\left[ \Delta
_{0,k}^{\ast }\Phi _{1,k}+\Delta _{0,k}\Phi _{1,k}^{\ast }\right] }{2},
\notag
\end{eqnarray}%
where $\Delta _{0,k}=\Delta _{0}\exp \left\{ \pm i\varphi /2+iUx/\xi
_{S}\right\} ,$ $\varphi $ is order parameter phase difference across the
barrier and the coefficient $U$ describes the linear growth of phase
difference due to the supercurrent in the electrodes. Corrections to $\Delta
_{0}$ and $\Phi _{S,k}$ are supposed to be small
\begin{equation}
\left\vert \Delta _{1,k}\right\vert \ll \Delta _{0},\quad \left\vert \Phi
_{1,k}\right\vert \ll \Delta _{0}.  \label{Ap_RC_Fi}
\end{equation}%
The approximation is valid if the right hand side of Eq.(\ref{GammM}) is
also small, so that
\begin{equation}
\xi _{S}\frac{\partial }{\partial x}\Phi _{1,k}=\Xi _{k}(\omega ),
\label{BCLin}
\end{equation}%
\begin{equation}
\Xi _{k}(\omega )=\pm \gamma _{M}\frac{\omega G_{F0,k}A_{0,k}}{\pi
T_{c}G_{0}^{2}}+\gamma \frac{\omega G_{F0,k}^{2}B_{k}}{\widetilde{\omega }%
_{k}G_{0}^{2}}  \label{sigotom}
\end{equation}%
\begin{equation*}
G_{F0,k}=\frac{\omega \vartheta _{k}}{\sqrt{\omega ^{2}\vartheta
_{k}^{2}+\Delta _{0}^{2}G_{0}^{2}}},
\end{equation*}%
where $\vartheta _{k}=\left( G_{0}+\widetilde{\omega }_{k}\gamma _{BM}/\pi
T_{c}\right) ,$ and $\left\vert \Xi (\omega )\right\vert \ll \Delta _{0}.$
From the structure of the linearized Usadel equations and the boundary
conditions (\ref{BCLin}) it follows that there are first order corrections
only to the magnitudes $\Theta $ and $\Delta _{1}$ of functions $\Phi _{1}$
and $\Delta _{1,k}$ respectively, while the phases of all of these functions
coincide with those of $\Delta _{0,k}.$ In this case
\begin{equation}
\tilde{\Phi}_{1,k}=\Theta \exp \left\{ \pm i\frac{\varphi }{2}\right\}
,\quad \Delta _{1,k}=\Delta _{1}\exp \left\{ \pm i\frac{\varphi }{2}\right\}
\label{formFi}
\end{equation}%
and due to the symmetry of the structure we have%
\begin{equation*}
\widetilde{\omega }_{R}=\widetilde{\omega }_{L}=\widetilde{\omega },\quad
G_{F0,k}=G_{F0},\quad \vartheta _{k}=\vartheta
\end{equation*}%
\begin{equation*}
\frac{A_{0,k}}{\Delta _{0}}=C_{0}\exp \left\{ \pm i\frac{\varphi }{2}%
\right\} ,\quad C_{0}=\frac{\widetilde{\omega }G_{0}}{\omega \vartheta },
\end{equation*}%
\begin{eqnarray}
\Xi _{k}(\omega ) &=&\frac{G_{F0}}{G_{0}\vartheta }\left[ \pm \gamma _{M}%
\frac{\widetilde{\omega }}{\pi T_{c}}\cos \frac{\varphi }{2}+\right.
\label{sigmaP} \\
&&\left. +i\left( \gamma _{M}\frac{\widetilde{\omega }}{\pi T_{c}}+2\frac{%
\gamma }{\gamma _{BI}}G_{F0}\right) \sin \frac{\varphi }{2}\right] .
\end{eqnarray}%
To write (\ref{sigmaP}), we also used the fact that in the first order with
respect to $\left\vert \Xi (\omega )\right\vert $ the magnitudes of
functions $\Phi _{S,k}$ in (\ref{GammaBM}) equal to $\Delta _{0}$ and that $%
G_{S}=G_{0}.$

Substituting\ (\ref{rigid}), (\ref{formFi}) into (\ref{EqUS}), (\ref{EqUSF}%
), we arrive at the following boundary problem for $\Theta $ and $\Delta
_{1} $
\begin{equation}
-\xi _{S}^{2}\frac{\pi T_{c}}{\sqrt{\omega ^{2}+\Delta _{0}^{2}}}\frac{%
\partial ^{2}}{\partial x^{2}}\Theta +\Theta =\Delta _{1},  \label{LinF}
\end{equation}%
\begin{equation}
\Delta _{1}\left[ \ln \frac{T}{T_{c}}+\pi T\sum_{\omega =-\infty }^{\infty }%
\frac{1}{|\omega |}\right] -\pi T\sum_{\omega =-\infty }^{\infty }\frac{%
\omega \Theta G_{0}}{(\omega ^{2}+\Delta _{0}^{2})}=0,  \label{LinD}
\end{equation}%
\begin{equation}
\xi _{S}\frac{\partial }{\partial x}\Theta (\pm d_{F})=\left[ \mathrm{Re}\Xi
_{k}(\omega )\cos \frac{\varphi }{2}\pm \mathrm{Im}\Xi _{k}(\omega )\sin
\frac{\varphi }{2}\right] ,  \label{BCmod}
\end{equation}%
\begin{equation}
\Theta (\pm \infty )=0.  \label{BCbulk}
\end{equation}%
Due to the symmetry of the problem it is enough to solve the equations (\ref%
{LinF})-(\ref{BCbulk}) only in one of the electrodes, namely, for $x\geq
d_{F}.$ Using the equation for $\Delta _{0}(T)$
\begin{equation}
\ln \frac{T}{T_{c}}+\pi T\sum_{\omega =-\infty }^{\infty }\frac{1}{|\omega |}%
=\pi T\sum_{\omega =-\infty }^{\infty }\frac{1}{\sqrt{\omega ^{2}+\Delta
_{0}^{2}}}  \label{delta0}
\end{equation}%
and the symmetry relation $\Theta (\omega )=\Theta (-\omega )$ we can
rewrite the selfconsistency equation in the form
\begin{equation}
\Delta _{1}\Sigma _{2}=\pi T\sum_{\omega >0}^{\infty }\frac{\pi T_{c}\omega
^{2}}{(\omega ^{2}+\Delta _{0}^{2})^{2}}\xi _{S}^{2}\frac{\partial ^{2}}{%
\partial x^{2}}\Theta  \label{LinD1}
\end{equation}%
\begin{equation}
\Sigma _{2}=\pi T\sum_{\omega >0}\frac{\Delta _{0}^{2}}{(\omega ^{2}+\Delta
_{0}^{2})^{3/2}}.  \label{sigma2}
\end{equation}%
The solution of (\ref{LinF}), (\ref{LinD1}) is%
\begin{equation}
\Delta _{1}=\sum_{\Omega >0}^{\infty }\delta _{\Omega }\exp (-q_{\Omega }%
\frac{x-d_{F}}{\xi _{S}}),  \label{Ap2_11Ma}
\end{equation}%
\begin{equation*}
\Theta =\sum_{\Omega >0}^{\infty }\frac{\delta _{\Omega }\sqrt{\omega
^{2}+\Delta _{0}^{2}}}{\sqrt{\omega ^{2}+\Delta _{0}^{2}}-\pi T_{c}q_{\Omega
}^{2}}\exp (-q_{\Omega }\frac{x-d_{F}}{\xi _{S}}),
\end{equation*}%
where the coefficients $\delta _{\Omega }$ and $q_{\Omega }$ satisfy the
equation
\begin{equation}
\Sigma _{2}=\pi T\sum_{\omega >0}^{\infty }\frac{\omega ^{2}}{(\omega
^{2}+\Delta _{0}^{2})^{3/2}}\frac{q_{\Omega }^{2}\pi T_{c}}{\sqrt{\omega
^{2}+\Delta _{0}^{2}}-\pi T_{c}q_{\Omega }^{2}},  \label{Eq_qq}
\end{equation}%
\begin{equation}
\sum_{\Omega >0}^{\infty }\frac{q_{\Omega }\delta _{\Omega }}{(\sqrt{\omega
^{2}+\Delta _{0}^{2}}-\pi T_{c}q_{\Omega }^{2})}=-\frac{\Delta _{0}P(\varphi
,\omega )}{\sqrt{\omega ^{2}+\Delta _{0}^{2}}}  \label{EqdeltaOm}
\end{equation}%
and $P(\varphi ,\omega )=\mathrm{Re}\Xi _{R}(\omega )\cos (\varphi /2)+%
\mathrm{Im}\Xi _{R}(\omega )\sin (\varphi /2).$ Multiplying Eq.(\ref%
{EqdeltaOm}) on $\omega ^{2}(\omega ^{2}+\Delta _{0}^{2})^{-3/2},$ summing
both sides of this equation on $\omega $ and making use of (\ref{Eq_qq}) one
can transform (\ref{EqdeltaOm}) into the system of equations for the
coefficients $\delta _{\Omega }$ which yield
\begin{equation}
\delta _{\Omega }=-\pi T\frac{\pi T_{c}\Delta _{0}\Omega ^{2}q_{\Omega }}{%
\Sigma _{2}(\Omega ^{2}+\Delta _{0}^{2})^{2}}\Lambda (\Omega ,\varphi ),
\label{solDelOm}
\end{equation}%
where%
\begin{equation*}
\Lambda (\Omega ,\varphi )=\left[ \gamma _{M}K_{1}(\Omega )+\frac{\gamma }{%
\gamma _{BI}}K_{2}(\Omega )(1-\cos \varphi )\right]
\end{equation*}

\begin{equation}
K_{1}(\Omega )=\frac{\Omega }{\pi T_{c}G_{0}}\sqrt{\frac{\sqrt{p^{2}+q^{2}}+p%
}{2(p^{2}+q^{2})}},  \label{K}
\end{equation}%
\begin{equation*}
K_{2}(\Omega )=\frac{pG_{0}+(Hq+p\Omega )\gamma _{BM}/\pi T_{c}}{%
G_{0}(p^{2}+q^{2})},
\end{equation*}%
\begin{equation}
q=2\gamma _{BM}\frac{H}{\pi T_{c}}(\gamma _{BM}\frac{\Omega }{\pi T_{c}}%
+G_{0}),  \label{qq}
\end{equation}%
\begin{equation}
p=1+\frac{\Omega ^{2}-H^{2}}{\left( \pi T_{c}\right) ^{2}}\gamma
_{BM}^{2}+2G_{0}\frac{\Omega }{\pi T_{c}}\gamma _{BM}.  \label{pp}
\end{equation}%
Here $\Omega =\pi T(2m+1)$ are the Matsubara frequencies$.$

As a result, the solution of the boundary problem (\ref{LinF})-(\ref{BCbulk}%
) has the form
\begin{equation}
\Delta _{1}=-\pi T\sum_{\Omega >0}\frac{\pi T_{c}\Delta _{0}\Omega
^{2}q_{\Omega }\exp (-q_{\Omega }\frac{x-d_{F}}{\xi _{S}})}{\Sigma
_{2}(\Omega ^{2}+\Delta _{0}^{2})^{2}}\Lambda (\Omega ,\varphi ),
\label{solpopD}
\end{equation}%
\begin{equation}
\Theta =-\pi T\sum_{\Omega >0}\frac{\pi T_{c}\Delta _{0}\Omega ^{2}q_{\Omega
}\Lambda (\Omega ,\varphi )\exp (-q_{\Omega }\frac{x-d_{F}}{\xi _{S}})}{%
\Sigma _{2}(\Omega ^{2}+\Delta _{0}^{2})^{2}(1-\pi T_{c}q_{\Omega
}^{2}G_{0}/\omega )}.  \label{SolpopF}
\end{equation}%
In particular, at $x=d_{F}$ from (\ref{solpopD}) and (\ref{SolpopF}) we have
\begin{equation}
\frac{\Theta (d_{F})}{\Delta _{0}}=-\gamma _{M}\Sigma _{F1}-\frac{\gamma }{%
\gamma _{BI}}\Sigma _{F2}(1-\cos \varphi ),  \label{F_dF}
\end{equation}%
\begin{equation}
\Sigma _{F1}=\pi T\sum_{\Omega >0}\frac{\pi T_{c}\Omega ^{2}q_{\Omega
}K_{1}(\Omega )}{\Sigma _{2}(\Omega ^{2}+\Delta _{0}^{2})^{2}(1-\pi
T_{c}q_{\Omega }^{2}G_{0}/\omega )},  \label{S_F1}
\end{equation}%
\begin{equation}
\Sigma _{F2}=\pi T\sum_{\Omega >0}\frac{\pi T_{c}\Omega ^{2}q_{\Omega
}K_{2}(\Omega )}{\Sigma _{2}(\Omega ^{2}+\Delta _{0}^{2})^{2}(1-\pi
T_{c}q_{\Omega }^{2}G_{0}/\omega )}.  \label{S_F2}
\end{equation}

To calculate the sums (\ref{S_F1}), (\ref{S_F2}) one needs to know the
expression for the coefficients $q_{\Omega }$ which can be in general
obtained from numerical solution of Eq.(\ref{Eq_qq}). Since the main
contribution to the sums (\ref{S_F1}), (\ref{S_F2}) comes from large $\Omega
,$ the asymptotic behavior of $q_{\Omega }$ at large $\Omega $ can be used%
\begin{equation}
q_{\Omega }^{2}=\alpha \frac{\sqrt{\Omega ^{2}+\Delta _{0}^{2}}}{\pi T_{c}}%
,\quad \alpha =1-\frac{\pi T^{2}}{\Omega T_{c}}\ln \frac{\sqrt{\Omega
^{2}+\Delta _{0}^{2}}}{\pi T}.  \label{q_as}
\end{equation}

The developed method is valid if the following condition is fulfilled%
\begin{equation}
(\gamma _{M}+\frac{\gamma }{\gamma _{BI}})\max \left\{ 1,\ \ln \left[ \frac{%
H^{2}+\left( \pi T_{c}\right) ^{2}}{\min \left\{ \gamma _{BM}^{2},\gamma
_{M}^{2}\right\} \left( \pi T\right) ^{2}}\right] \right\} \ll 1,
\label{cond_R}
\end{equation}%
\begin{equation*}
\gamma _{B}\ll \gamma _{BI}.
\end{equation*}

Therefore for the function $\Phi _{S,k}$ in Eq.(\ref{Arov}) we get%
\begin{equation}
\Phi _{S,k}=(\Delta _{0}+\Theta (d_{F}))\exp \left\{ \mp i\varphi /2\right\}
,  \label{Ficorr}
\end{equation}%
and substituting (\ref{Ficorr}) into (\ref{GammaBM}) we finally obtain

\begin{eqnarray}
A_{k} &=&\left[ \Delta _{0}+\frac{\omega \mu C_{0}}{\widetilde{\omega }}%
\Theta (d_{F})\right] C_{0}\exp \left\{ \pm i\varphi /2\right\} \mp
\label{ApF} \\
&&\mp 2i\frac{\gamma _{B}}{\gamma _{BI}}\frac{\widetilde{\omega }%
G_{0}G_{F0}\Delta _{0}}{\omega \vartheta ^{2}}\sin \frac{\varphi }{2},
\end{eqnarray}%
\begin{equation*}
\mu =1+G_{0}\widetilde{\omega }\gamma _{BM}/\pi T_{c}
\end{equation*}%
From the structure of coefficients $\tilde{A}_{R,L}$ we see that the
corrections to the supercurrent across the SFIFS tunnel junction
leads not only to the reduction of the critical current of the
structure, but also to changes in the J$_{s}(\varphi )$ relation.

\textbf{The J}$_{S}(\varphi )$\textbf{\ relation.} Using the the standard
expression for the supercurrent \cite{Golubov1}, the boundary condition (\ref%
{BI_2}) and Eq.(\ref{ApF}) we can write down the supercurrent $I$\textit{\ }%
across the SFIFS junction in the form%
\begin{equation}
I==(J_{0}+J_{11})\sin \varphi +J_{12}\sin 2\varphi ,  \label{curTfi}
\end{equation}%
where
\begin{equation}
J_{0}=\frac{\pi T}{eR_{N}}\sum_{\omega =-\infty }^{\infty }\frac{\Delta
_{0}^{2}C_{0}^{2}}{\widetilde{\omega }^{2}+C_{0}^{2}\Delta _{0}^{2}},\quad
C_{0}=\frac{\widetilde{\omega }G_{0}}{\omega \vartheta }  \label{J0p}
\end{equation}%
\begin{equation*}
J_{11}=-\frac{2\pi T}{eR_{N}}\sum_{\omega =-\infty }^{\infty }\frac{\Delta
_{0}^{2}C_{0}^{2}}{(\widetilde{\omega }^{2}+C_{0}^{2}\Delta _{0}^{2})^{2}}%
\left[ \gamma _{M}\frac{\widetilde{\omega }\omega C_{0}\mu }{\Delta _{0}}%
\Sigma _{F1}+\right.
\end{equation*}%
\begin{equation}
\left. +\frac{\gamma _{B}}{\gamma _{BI}}\frac{\widetilde{\omega }^{2}G_{F0}}{%
\vartheta }+\frac{\gamma }{\gamma _{BI}}\frac{\widetilde{\omega }\omega
C_{0}\mu }{\Delta _{0}}\Sigma _{F2}\right] ,  \label{j11_f}
\end{equation}%
\begin{eqnarray}
J_{12} &=&-\frac{\pi T}{eR_{N}}\sum_{\omega =-\infty }^{\infty }\frac{\Delta
_{0}^{2}C_{0}^{3}}{(\widetilde{\omega }^{2}+C_{0}^{2}\Delta _{0}^{2})^{2}}%
\left[ \frac{\gamma _{B}}{\gamma _{BI}}\frac{G_{F0}\Delta _{0}^{2}C_{0}}{%
\vartheta }-\right.  \label{J12_f} \\
&&\left. -\frac{\gamma }{\gamma _{BI}}\frac{\widetilde{\omega }\omega \mu
\Sigma _{F2}}{\Delta _{0}}\right]
\end{eqnarray}

Expression (\ref{J0p}) has been obtained previously in \cite{VolkovAF}-\cite%
{Golubov1}. The $\varphi -$independent correction to it, $J_{11},$ is
negative and describes the suppression of $\sin \varphi $ component of the
supercurrent. The first term in Eq.(\ref{j11_f}) proportional to $\gamma
_{M} $ takes into account the suppression of superconductivity in S
electrodes due to proximity with thin F layer. The last two terms
proportional to $\gamma _{BI}^{-1}$ describe the suppression of
superconductivity by the current across the junction. The larger $\gamma
_{B} $ and $\gamma $ the weaker is the superconductivity induced into F
layer and the stronger is the influence of this effect.

The sign of the second harmonic $J_{12}$ depends on the relation between $%
\gamma _{B}$ and $\gamma .$ At $\gamma _{B}=0$ it is positive and
$J(\varphi )$ relation (\ref{curTfi}) has a maximum at $\varphi
=\varphi _{\max }<\pi /2$ . Such a shift was predicted earlier near
$T_{c}$ for SIS tunnel junctions and is due to the suppression of
superconductivity near the barrier by a supercurrent \cite{kupr1}.
Increase of $\gamma _{B}$ leads to additional phase shifts at both
SF interfaces\ and provides the mechanism for the shift of the
$\varphi _{\max }$ into the region $\varphi >\pi /2$. As a result,
at sufficiently large $\gamma _{B}$ the amplitude $J_{12}$ changes
its sign and $\varphi _{\max }$ shifts to $\varphi >\pi /2.$ Such a
competition between suppression by a supercurrent and by proximity
effect
was first analyzed in the SNS\ junctions\cite{SNSTc} at $T\approx T_{c}$%
.This fact is in the full agreement with the results of numerical
calculations summarized in \cite{GKI}.

The physical reason for different signs of $J_{12}$ can be easily understood
if we consider the two cases separately. Suppose first that $\gamma _{B}$ is
finite. In this case the SFIFS\ structure may be considered as a system of
three Josephson junctions in series as shown schematically in Fig.1. For
rough estimates one can assume that the phase $\chi $ of $\Phi _{F,k}$ does
not depend on $\omega .$ Demanding the equality of the currents across FIF\
and FS interfaces and taking into account that $I_{C}\propto \gamma
_{BI}^{-1}\ll I_{C1}\propto \gamma _{B}^{-1}$ for $\chi $ we will have%
\begin{equation*}
\chi =\varphi /2-\frac{I_{C}}{I_{C1}}\sin 2\chi .
\end{equation*}%
Substituting this $\chi $ into the expression for the supercurrent across
FIF\ interface, we get%
\begin{equation}
I=I_{C}\sin (\varphi -\frac{I_{C}}{I_{C1}}\sin \varphi )\approx I_{C}(\sin
\varphi -\frac{\gamma _{B}}{\gamma _{BI}}\sin 2\varphi ).  \label{ifi1}
\end{equation}%
Therefore with increasing $\gamma _{B}$ the phase partly jumps at the FS
interfaces leading to a continuous crossover from the Josephson effect
lumped at $x=0$ to the phase drop distributed at $\left\vert x\right\vert
\leq d_{F}$. In a full agreement with the theory of double barrier devices
\cite{GKI} this crossover results in appearance of second harmonic in $%
J_{S}(\varphi )$ with negative sign which provides maximum $J_{S}(\varphi )$
achieved at $\varphi \geq \pi /2.$

\begin{figure}[tbp]
\includegraphics[width=2.8in ]{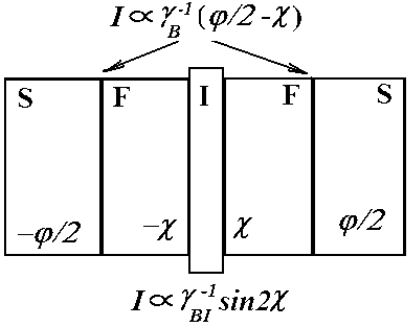}
\caption{The phase distribution in a SFIFS junction}
\end{figure}

If $\gamma _{B}=0,$ the structure is always lumped at $x=0$ and the main
effect is the suppression of superconductivity by supercurrent in the
vicinity of FIF\ interface as shown schematically in Fig.2. The resulting
contribution to the full current is%
\begin{equation}
I_{\omega }\propto \gamma _{BI}^{-1}(\Delta _{0}-\xi _{S}\frac{\partial
\Theta }{\partial x})\sin \varphi \propto \frac{\Delta _{0}}{\gamma _{BI}}(1-%
\frac{\sin ^{2}\frac{\varphi }{2}}{\gamma _{BI}})\sin \varphi .  \label{ifi2}
\end{equation}%
It follows directly from (\ref{ifi2}) that the amplitude of the second
harmonic is positive.

The competition of the above two mechanisms of $I(\varphi )$ deformation is
clearly seen from Eq. (\ref{J12_f}).

The general expressions (\ref{J0p})-(\ref{J12_f}) can be simplified in
several limiting cases.

\begin{figure}[tbp]
\includegraphics[width=2.8in ]{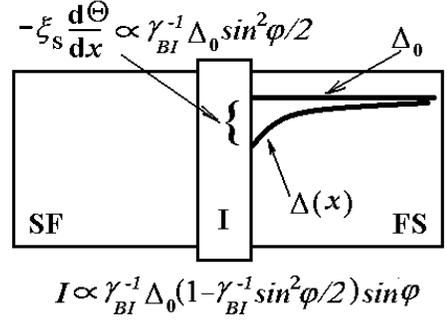}
\caption{Depairing by current near the tunnel barrier}
\end{figure}

In the symmetric SNINS tunnel junctions $H=0$ in both electrodes and in the
first approximation from (\ref{J0p}) the earlier result from \cite{GKJ} is
reproduced%
\begin{equation}
J_{0}=\frac{2\pi T}{eR_{N}}\sum_{\omega \geq 0}^{\infty }\frac{\Delta
_{0}^{2}}{(\omega ^{2}+\Delta _{0}^{2})\Theta (\omega )},  \label{J0SNINS}
\end{equation}%
\begin{equation*}
\Theta (\omega )=(1+2G_{0}\omega \gamma _{BM}/\pi T_{c}+(\omega \gamma
_{BM}/\pi T_{c})^{2})
\end{equation*}%
while (\ref{j11_f}) and (\ref{J12_f}) reduce to%
\begin{equation}
J_{11}=-\frac{4\pi T}{eR_{N}}\left[ \gamma _{M}\Sigma _{4}+\frac{\gamma _{B}%
}{\gamma _{BI}}\Sigma _{5}+\frac{\gamma }{\gamma _{BI}}\Sigma _{6}\right] ,
\label{j11_snins}
\end{equation}%
\begin{equation}
J_{12}=-\frac{2\pi T}{eR_{N}}\left[ \frac{\gamma _{B}}{\gamma _{BI}}\Sigma
_{7}-\frac{\gamma }{\gamma _{BI}}\Sigma _{6}\right] ,  \label{j12_snins}
\end{equation}%
where%
\begin{equation}
\Sigma _{4}=\sum_{\omega >0}^{\infty }\frac{\Delta _{0}G_{0}\vartheta \mu
\Sigma _{F1}}{(\omega ^{2}+\Delta _{0}^{2})\Theta ^{2}(\omega )},  \label{s4}
\end{equation}%
\begin{equation}
\Sigma _{5}=\sum_{\omega >0}^{\infty }\frac{\Delta _{0}^{2}\vartheta ^{2}}{%
(\omega ^{2}+\Delta _{0}^{2})\Theta ^{5/2}(\omega )},  \label{s5}
\end{equation}%
\begin{equation}
\Sigma _{6}=\sum_{\omega >0}^{\infty }\frac{G_{0}\Delta _{0}\vartheta \mu
\Sigma _{F2}}{(\omega ^{2}+\Delta _{0}^{2})\Theta ^{2}(\omega )},  \label{s6}
\end{equation}%
\begin{equation}
\Sigma _{7}=\sum_{\omega >0}^{\infty }\frac{\Delta _{0}^{4}}{(\omega
^{2}+\Delta _{0}^{2})^{2}\Theta ^{5/2}(\omega )},  \label{s7}
\end{equation}%
and $G_{0}=\omega /\sqrt{\omega ^{2}+\Delta _{0}^{2}}.$

In the limit $\gamma \rightarrow 1,$ $H,\gamma _{M},\gamma _{B},\gamma
_{BM}\rightarrow 0$ the SFIFS structure transforms into SIS tunnel junction.
In this case
\begin{equation}
C_{0}=1,\quad A_{pR,L}=\left[ \Delta _{0}+\Theta (d_{F})\right] \exp \left\{
\pm i\varphi /2\right\} ,  \label{CSIS}
\end{equation}%
\begin{equation}
\Theta (d_{F})=-\frac{2}{\gamma _{BI}}\pi T\sum_{\Omega >0}\frac{\pi
T_{c}\Delta _{0}\Omega ^{2}q_{\Omega }\sin ^{2}\frac{\varphi }{2}}{\Sigma
_{2}(\Omega ^{2}+\Delta _{0}^{2})^{2}(1-\pi T_{c}q_{\Omega }^{2}G_{0}/\omega
)},  \label{F1SIS}
\end{equation}%
and for the supercurrent $I$ in the first approximation we have the well
known result of Ambegakaokar-Baratoff theory \cite{AB}
\begin{equation}
I=\frac{2\pi T}{eR_{N}}\sum_{\omega >0}^{\infty }\frac{\Delta _{0}^{2}}{%
\omega ^{2}+\Delta _{0}^{2}}\sin \varphi .  \label{J0SIS}
\end{equation}%
Using (\ref{Eq_qq}) for $J_{11}$ and $J_{12}$ it is easy to get
\begin{equation}
J_{11}=-\frac{\Delta _{0}}{eR_{N}}2\Sigma _{3},\ J_{12}=\frac{\Delta _{0}}{%
eR_{N}}\Sigma _{3},  \label{J1SIS}
\end{equation}%
\begin{equation}
\Sigma _{3}=\frac{4}{\gamma _{BI}}\pi T\sum_{\Omega >0}\frac{\Delta
_{0}\Omega ^{2}}{(\Omega ^{2}+\Delta _{0}^{2})^{2}q_{\Omega }},
\label{J12SIS}
\end{equation}%
and the full current across the tunnel junctions is
\begin{equation}
I=\frac{\Delta _{0}}{eR_{N}}\left[ \frac{\pi }{2}\tanh \frac{\Delta _{0}}{2T}%
-2\Sigma _{3}\right] \sin \varphi +\frac{\Delta _{0}\Sigma _{3}}{eR_{N}}\sin
2\varphi .  \label{FullSIS}
\end{equation}%
The critical current achieves at phase difference $\varphi _{c}$%
\begin{equation}
\varphi _{c}=\frac{\pi }{2}-\frac{4\Sigma _{3}}{\pi }\tanh ^{-1}\frac{\Delta
_{0}}{2T},  \label{ficrit}
\end{equation}%
and equals to%
\begin{equation*}
I_{c}\approx \frac{\Delta _{0}}{eR_{N}}\left[ \frac{\pi }{2}\tanh \frac{%
\Delta _{0}}{2T}-2\Sigma _{3}\right]
\end{equation*}%
and at $T\rightarrow 0$ the $I(\varphi )$ simplifies to
\begin{equation}
I_{c}\approx \left[ \frac{\Delta _{0}}{eR_{N}}\frac{\pi }{2}-\frac{1.92}{%
\gamma _{BI}}\left( \frac{\pi T_{c}}{\Delta _{0}}\right) ^{3/2}\right] .
\end{equation}%
At $T\approx T_{c}$ Eqs.(\ref{J1SIS}) transform to the result obtained in
\cite{kup}.

\textbf{Conclusions.} In summary, we have studied the current-phase
relations $J_{S}(\varphi )$ in SFIFS, SNINS and SIS junctions in the regime
when the second harmonic of $J_{S}(\varphi )$ is not small. To solve this
problem selfconsistently, we have developed the analytical method for
solving the linearized Usadel equations. This solution describes a weak
suppression of superconducting state in a superconductor caused either by
proximity with normal or ferromagnetic material or by a current in composite
SN or SF proximity systems. The method is rather general and can be applied
to a wide spectrum of proximity problems.

We have demonstrate that the full current across the structure (\ref{curTfi}%
) consists of the sum $\sin \varphi $ and $\sin 2\varphi $ components and
have calculated the amplitudes $(J_{0}+J_{11})$ and $J_{12}$ of these
components. In SIS and SNINS structures the corrections $J_{11}$ and $J_{12}$
to the previously calculated critical current $J_{0}$ are small. The $%
J(\varphi )$ curve is slightly deformed so that the maximum value of the
supercurrent achieved at phase difference $\varphi _{c}$ which can be
smaller or larger $\pi /2$ for positive and negative sign of $J_{12}$
respectively. In SFIFS junctions $J_{0}=0$ at the point of the transition
from $"0"$ to $"\pi "$ state. It means that in this case the calculated
values $J_{11}$ and $J_{12}$ determine the $J(\varphi )$ curve. Since the
amplitudes $J_{11}$ and $J_{12}$ may have comparable magnitude, the $%
J(\varphi )$ measured experimentally can be essentially different from $\sin
\varphi $. The validity of the developed approach is determined by
inequalities (\ref{cond_R}) and $\gamma _{B}\ll \gamma _{BI}.$ These
conditions also determine the validity of rigid boundary conditions in the
models\cite{GKI} describing the properties of SFIFS, SNINS and SIS tunnel
junctions.

This work has been supported in part by Russian Ministry of Education and
Science, RFBR Grant N 04 0217397-a, INTAS Grant 01- 0809, NWO-RFBR
cooperation programme 047.016.005 and ESF PiShift programme.

\end{document}